\documentclass[12pt,preprint]{aastex}
\slugcomment{Invited review talk to appear in the proccedings of the
meeting ``New Quests in Stellar Astrophysics: the Link between Stars
and Cosmology'', M. Chavez, A. Bressan, A. Buzzoni \& D. Mayya, eds, Kluwer
Academic Publishers}

\shorttitle{Evolution of Early Type Galaxies} 

\shortauthors{L\'opez-Cruz et al.}

\shorttitle{Early Type Galaxies in Clusters}

\shortauthors{L\'opez-Cruz et al.}

\begin{document}

\title{Excursions into the Evolution of Early-Type Galaxies in
 Clusters}

\author{Omar L\'opez-Cruz\altaffilmark{1}}
\affil{Instituto Nacional de Astrof{\'\i}sica, Optica y
Electr\'onica (INAOE)  \&  DAUG , M\'exico}

\author{David Schade}
\affil{CADC, Herzberg Institute of Astrophysics (HIA),  Canada}

\author{L. Felipe Barrientos}
\affil{Departamento de Astron. \&  Astrof{\'\i}s. Pontificia Universidad
Cat\'olica de Chile (PUC), Chile}

\author {Michael D. Gladders\altaffilmark{2} \& H.~K.~C. Yee} 
\affil{Department of Astronomy and Astrophysics,
University of Toronto (U of T), Canada}

\author{Tadayuki Kodama}
\affil{Department of Astronomy, University of Tokyo, Japan}

\altaffiltext{1}{Fulbright Visiting  Scholar at the  Observatories of the
Carnegie Institution of Washington (OCIW)}

\altaffiltext{2}{Carnegie Fellow, OCIW}

\begin{abstract}
  Recent observations have revealed that early-type galaxies (ETG)  in
  clusters comprise an old galaxy population that is evolving
  passively. We review some recent  observations from the
  ground and from the HST that show that ETG have
  undergone a significant amount of luminosity evolution. This
  evolution is traced by two  projections of the fundamental
  plane (FP): the size-magnitude relation (SMR) and the
   color-magnitude relation (CMR).  We will briefly discuss the
  relevance of all these results in the context of the universality of
  the IMF.  

\end{abstract}


\section{Introduction}  
        
The opening line of Walter Baade's lectures on the evolution of stars
and galaxies \citep{Ba63} reads: ``The study of the evolution of stars
and galaxies is in a stage of rapid development, and we are still at
the beginning''. Almost half a century later the former statement is
still timely,  as advances in technology have opened new windows for
the exploration of galaxy evolution.  The first clue that galaxies
change through cosmic time is that galaxies at high redshift ($z$)
appear systematically brighter than those at lower $z$. We may
identify two reasons for this brightening: it could be that either the
stellar content was intrinsically brighter at an earlier epoch or that
galaxies appear brighter due to the properties of the space-time.  We
are fortunate that those issues happened to be the main running themes
of this rather interesting conference. Clearly, if we want to
understand the brightening of high-$z$ galaxies, we need to understand
the role of stellar evolution in galaxies as well as the propagation
of light in different cosmological models. With these elements in mind let
us explore the evolution of galaxy cluster ETG and its possible
connections with the evolution of field ETG.  We note
that ETG were once considered the simplest stellar
systems. Nevertheless, ETG, despite their apparently simple appearance,
pose some difficult problems to models of galaxy formation and
evolution. Perhaps the most remarkable  is the inexistence of a
comprehensive theory of stellar formation in dynamically hot systems
\citep{Silk99}.

Clusters of galaxies provide us with probes to study the effects of
the environment on the evolution of galaxies
\citep{Oemler74,Dress84}. Indeed, it has been found that clusters
provide a hostile environment for some galaxies (e.g., dwarf
galaxies: \citealt{Lop97}; \citealt{Hil98}, and low surface brightness:
\citealt{GreWest98}). In contrast, ETG flourish in such rich
environments: this is just a paraphrase of the density-morphology
relation \citet{Dress80}.  Interestingly, the physical parameters of
these galaxies, such as colors, size (e.g., the effective radius $R$),
luminosity ($L$), velocity dispersion ($\sigma$), surface brightness
($I$), etc., have the same pairwise correlations originally found for
field ETG (e.g., $L\propto\sigma^{4}$).  The correlations among these
parameters are synthesized by the fundamental plane
($R\propto\sigma^{1.4\pm0.15}I^{-0.9\pm0.1}$; FP;
\citealt{DjorDa87}). Nevertheless, the effects of the environment may be
indicated by an augmented tightness in the correlations for cluster
galaxies \citep[e.g.,][]{Schade99,TFD01}.

In this review we present the results of some apparently disconnected
evolutionary studies. We have explored two projections of the
fundamental plane. These studies have led us to conclude that ETG in
the central regions (the inner $\approx 1\,h^{-1}_{50}\, {\rm
Mpc}$)\footnote{A Hubble constant of $H_{\circ}=50\,h_{50}\, {\rm
Mpc}\,{\rm km}^{-1}\,{\rm s}^{-1}$ is used throughout.}  of rich
clusters conform to an old population that has been evolving passively
since $z \sim 3$. We do not intend to present a comprehensive review
\citep[see elsewhere, e.g.,][]{KorDj89,BRS95,AdCS96,CeCa98,Ellis01};
however, we do intend to arrive at a unifying view on the evolution of
ETG.

\section{The Size-Magnitude Relation}

The work of \citet{BSLC96} was the first study from a series that
compared ETG in an intermediate-$z$ cluster (CL $0939 +4713$ at
$z=0.41$, observed with the HST) with those of a nearby rich cluster
(the Coma cluster at z=0.023, observed with the KPNO 0.9m) in a
bidimensional space that is defined by axes: size versus integrated
magnitude.  The size is defined as the core radius in a bidimensional
de Vacouleurs fit to the surface brightness distribution, and the
integrated magnitude was computed by the integration of the best
fit. The resulting size-magnitude relation (SMR) is expected as it
conforms one of the projections of the FP \citep{DjorDa87}.  When the
k-corrected SMR of Coma and CL $0939 +4713$ are compared, an offset
of $\Delta M_{B}=0.72 \pm 0.24$ is detected. This
offset indicates that the galaxies in CL $0939 +4713$ are brighter
than the ones in Coma. We can explain this result either by
assuming that the size remained fixed and the change was due to the
evolution of the underlying stellar population (passive evolution) or
by assuming that both the size and the magnitude were changing due to
mergers. Although mergers could have been important at the time of
cluster formation \citep{Sch01}; the merger hypothesis is not favored
because in the central regions of rich clusters the large cluster
velocity dispersions ($\sigma_{cluster}\geq 750\;\,{\rm km}\,{\rm
s}^{-1}$) would make merging inefficient. Moreover, the general {\em
merger rate}\footnote{For example {\em merger rate}
$\propto(1+z)^{2.3\pm0.7}$ \citep{Patt02}} is very low at $z<1$. Hence,
we can conclude that the observed changes are due to passive evolution.
Similar amount of luminosity evolution was reported by previous
observations in clusters around quasars \citep{YG87}.

Changes in luminosity can easily be explained with the aid of the
single burst model \citep[cf.,][]{T80,Buzz95,vDokkum98} where a burst
of star formation occurs at an epoch $t=t_{\circ}$. It can be shown
that the expected luminosity evolution is described by a power law
$L(t)\propto\left[\frac{1}{t-t_{\circ}}\right]^{\kappa}$, with
$\kappa=[1.3-0.27(s-1)]>0$ where $x=(s-1)$ is the slope of the IMF; it
has been found that $\kappa$ depends on the metallicity and the
passband\footnote{ A useful approximation valid at low $z$ is provided
by mpty universe model ($\Omega_{\circ}=0$). A simple relation results
for the luminosity evolution as a function of $z$: $L(z) \propto
\left[\frac{(1+z)(1+z_{\circ})H_{\circ}}{(z_{\circ}-z)}\right]^{\kappa}$,
where $z_{0}[\approx\,3]\,>z$ is the galaxy formation
redshift.}. Hence, as $t$ becomes larger than $t_{\circ}$ the
luminosity decreases, in agreement with the expected dimming in
luminosity as stars evolve off the giant branch.  Further studies
\citep{Schade96,Schade97} have included more clusters and a
larger number of galaxies and have confirmed and extended the early
result of \citeauthor{BSLC96}. Figure 1 taken from \citet{Schade97}
summarizes the results from our previous studies. It is found that
$\Delta M\propto z$, in addition the direct comparison with the models
of \citet{Buzz95} has given us a constraint on $s$.  The four most
deviant points in this figure are produced by poor
sampling. Therefore, we can safely conclude that our observations are
in agreement with models with $s=2.35$, i.e., $x=1.35$ the Salpeter
IMF \citep{Sal55}.


\begin{figure}[t]

\plotone{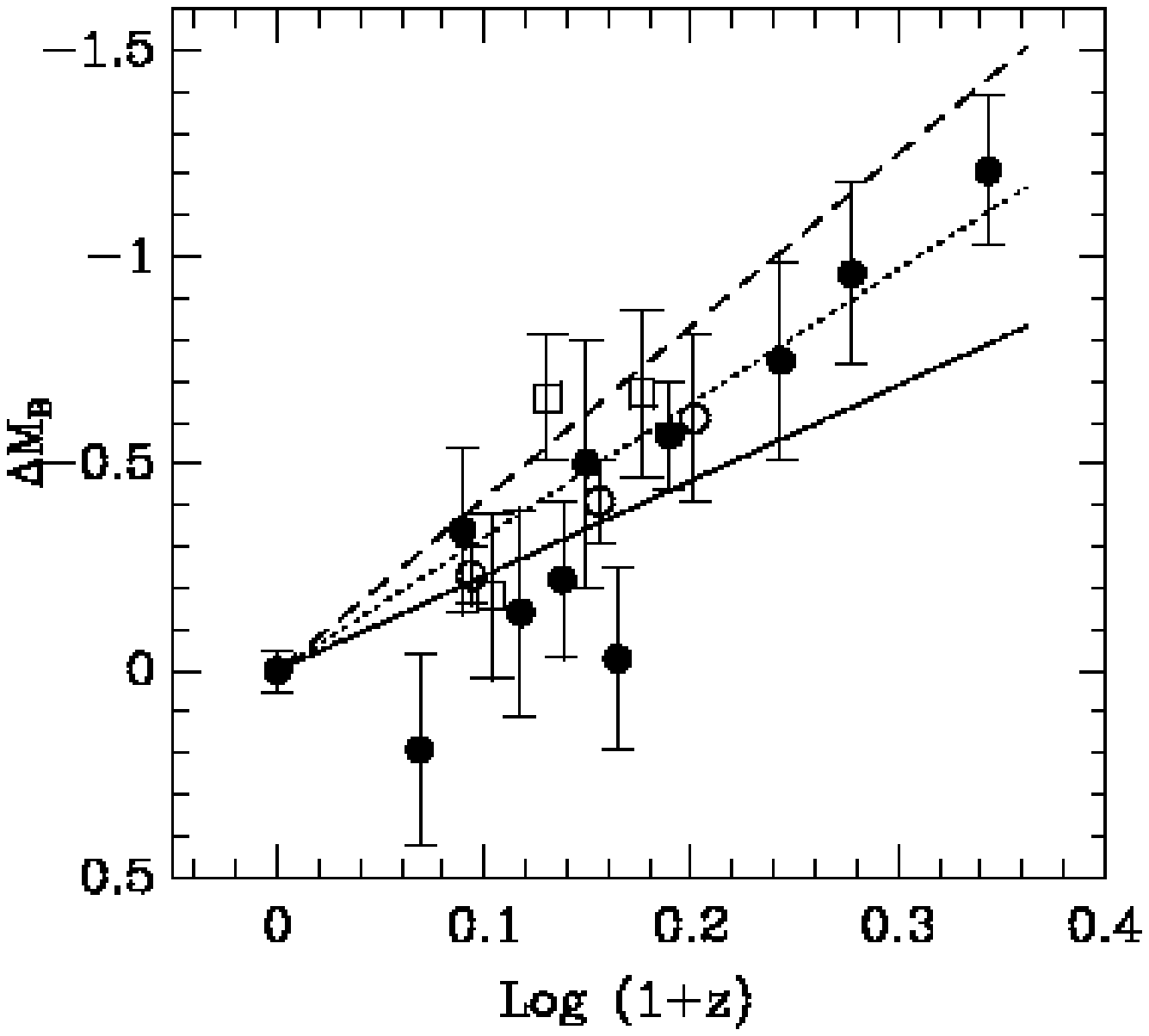}

\caption{ The luminosity shift $\Delta M_B$ including revised results
from Schade et al. (1996a) and from Barrientos, Schade, \&
L\'opez-Cruz (1996). Filled symbols are for clusters in the present
study, open circles are cluster points from Schade et al. (1996b)
(including 2 clusters in common that have been offset in redshift
slightly for clarity) and open squares are field galaxies from Schade
et al. (1996a). The lines are from the models of Buzzoni (1995) for
IMF power-law indices of $s=3.35$ (solid line), $s=2.35$ (short
dashed), and $s=1.35$ (long dashed). The models assume a present-day
age of 15 Gyr and we show the theoretical $\Delta M(V)$ whereas the
data is in $M_B$ where the evolution would be $\sim 0.15$ mag larger
by $z=1$. \label{figure1}}
\end{figure}

\section{The Color-Magnitude Relation}

The first pairwise relationship discovered among the physical
properties of ETG was the color-magnitude relation (CMR). It was
originally discovered by \citet{Baum59} as a trend in which the
brightest galaxies are also the reddest. In the same paper
\citeauthor{Baum59}, with the aid of a simple population synthesis
model, demonstrated that ETG were dominated by old {\em Population I}
stars, contrary to the then popular belief that globular clusters and
ETG had the same stellar make up.  \citeauthor{Baum59}'s observations
and their interpretation signified a turning point in our
understanding of ETG.  The work of \citet{San72}, \citet{Faber73}, and
\citet{VisSan77} showed that the CMR can be parameterized by a
straight line within a luminosity interval of about 8 magnitudes. This
remarkable property suggests that ETG from giants to dwarves, with the
exception of some brightest cluster galaxies (BCG)\footnote{Sometimes
BCG in cooling flow clusters show very blue cores
\citep[e.g.,][]{McO92,Lop97a} hence, some BCG deviate from the CMR},
have shared a common history of star formation, even when dwarfs and
giants are structurally and dynamically different: a dwarf elliptical
galaxy is not a scaled-down giant ETG \citep{FB94}. This difference in
dynamical behavior is further suggested by the break away of dwarves
from the FP and the SMR. The other important property is the
universality of the CMR at least for clusters at low redshift
\citep[e.g.,][]{SanVis78,BLE92,LCY02}. The CMR was first explained by
\citet{AY87} in terms of the combined effects of dissipative galaxy
formation and metallicity.

Three constraints on the last epoch of strong galaxy formation in ETG
can be derived from the study of the CMR.  \citet{BLE92} showed that
the dispersion of the CMR is a good indicator of the age spread in the
stellar population, since newer episodes would increase the dispersion
about the CMR.  We have found that the dispersion is smaller than the
limit of our observations for galaxies in the central regions of
clusters at lower redshift \citep[the LOCOS
sample,][]{Lop97a,Lop01}. The second constraint is provided by the
absolute color evolution of ETG, this can be measured by changes of
the zero point of the CMR \citep[see,][]{SED98}.  The third constraint
is provided by the slope of the CMR: in the classical scenarios
\citep[e.g.,][]{KA97} star formation is regulated by supernova driven
winds. In these models the strongest changes in the stellar population
properties occur during the first 5 Gyrs after formation. Hence if we
are able to identify evolution in the slope of the CMR, we could in
principle constrain the epoch of the last major starburst.  We have
attempted such a test using archive HST observations and LOCOS data
\citep{Gla98}. The merit of this test is that the slope, being a ratio
of colors, is independent of the calibrations and therefore we are
provided by an unbiased estimator that can be readily compared with
the models.

\begin{figure}[t]
\plotone{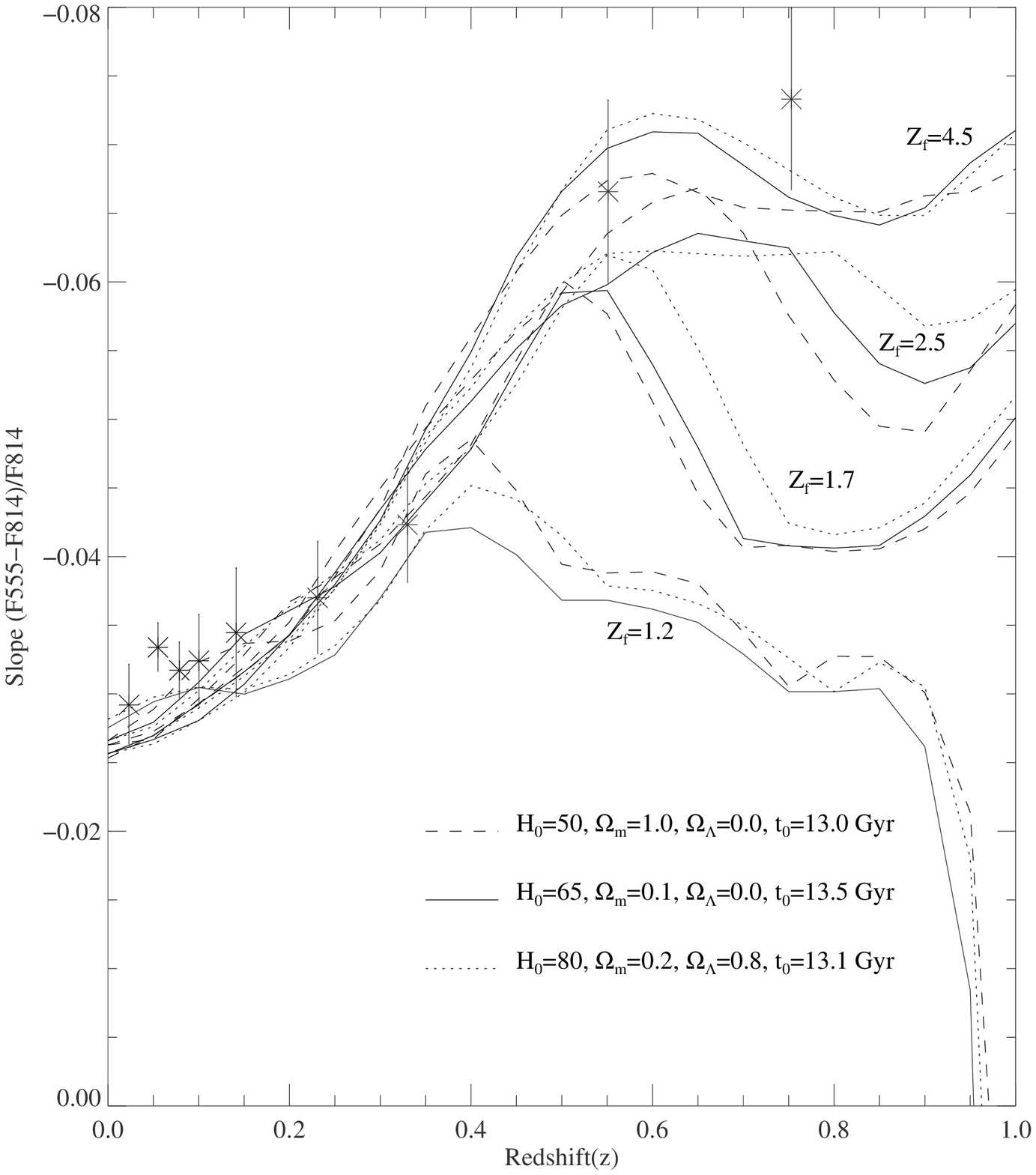}
\caption{ This figure has been taken from Gladders et al. (1998). The
predicted slopes for formation models (Kodama 1997, Kodama \& Arimoto
1997) using three different cosmologies to map age onto redshift.
Note that while the details of the turn-over in the slope for a model
at a given formation redshift changes slightly in different
cosmologies, the change is not significant enough to affect the
conclusion that at least some clusters form at $z>$2.
\label{figure2}}
\end{figure}

 The comparison of our observed CMR ~slopes to a suite of four models
 with different ~formation epochs and ~cosmologies is presented in
 Figure 2. The overall slope of the red sequence seems to be
 consistent with passive evolution for all clusters, implying that the
 populations we have sampled have similar evolutionary histories and
 time-scales. The observed slopes of the higher redshift clusters are
 perfectly consistent with a high formation redshift, and a
 conservative limit of $z\geq2$ can be set. Note that up to redshifts
 of z$\approx$0.4, observations of the slope of the red sequence have
 little power to distinguish between formation epochs with z$\geq$1.0.
 The change in slope at z$\leq$0.4 is exclusively due to blue-shifting
 of the rest-frame band-passes. However, beyond z$\approx$0.5, the
 expected slopes for the lowest-formation-epoch models rapidly diverge
 from the observed slopes. This turnover is a result of the differing
 metalicities in the ETG population as a function of mass; at ages
 younger than $\sim$ 5 Gyr, the color evolution of the stellar
 population in a metal-rich ETG is significantly different from that
 of a metal-poorer elliptical \citep{K97,KA97}. From Figure 2 it is
 evident that that the slope of the CMR provides a good marker on the
 last episode of star formation in cluster ETG that is almost
 independent of the cosmology.

\section{Discussion \& Conclusions}

The two studies that have been presented here strongly suggest that
ETG in clusters are evolving passively.  Also \citet{Pa96} and
\citet{LS01} using the Tolman test have shown indirect
evidence that also suggest that ETG are evolving passively.

We have used the  poor-man's approach to the FP in order to address the
evolution of ETG. Can we expect to detect the same  kind of evolution using
the FP itself? The answer is yes: there are many FP studies for
cluster galaxies at low and intermediate $z$ that have agreed that ETG
are evolving passively (e.g., \citet{JFK96}; \citet{vDokkum98}).

Figure 1 also reveals that the best agreement between the models and
the observations is reached with a Salpeter IMF. This is remarkable
since \citet{Sal55} defined the IMF for the solar
neighborhood. Field ETG at higher redshifts also reveal a
Salpeter IMF \citep{Schade99}. Star formation being such a complex
physical process has all the characteristic of self-organized
criticality \citep[see,][]{MS00}.  Our results along with
those  of \citeauthor{MS00} further supports the universality of the IMF.

Our general conclusion, which also agrees with many other studies 
\citep[see,][]{Pee02}, is that cluster ETG  coevally formed at
$z>3$ and have been evolving passively since then. Moreover, a single
burst with a Salpeter IMF suffices to explain the observations
presented in this review.


\acknowledgements
OLC research is supported in part by CONACyT-M\'exico Young
Researcher Grant No. J32098-E

\end {document}